\documentclass[twocolumn]{aastex631}

\usepackage{url}
\usepackage{afterpage}
\usepackage{lipsum}
\usepackage{enumitem}
\usepackage{physics}
\usepackage{multirow}
\usepackage{textgreek}
\usepackage{booktabs}
\usepackage{xfrac}
\hypersetup{pdfnewwindow=true,
     colorlinks=true,
     linkcolor=blue, 
     citecolor=blue,
	 final=true}
	 
\usepackage{etoolbox}
\makeatletter
\patchcmd{\NAT@citex}
  {\@citea\NAT@hyper@{%
     \NAT@nmfmt{\NAT@nm}%
     \hyper@natlinkbreak{\NAT@aysep\NAT@spacechar}{\@citeb\@extra@b@citeb}%
     \NAT@date}}
  {\@citea\NAT@nmfmt{\NAT@nm}%
   \NAT@aysep\NAT@spacechar\NAT@hyper@{\NAT@date}}{}{}
\patchcmd{\NAT@citex}
  {\@citea\NAT@hyper@{%
     \NAT@nmfmt{\NAT@nm}%
     \hyper@natlinkbreak{\NAT@spacechar\NAT@@open\if*#1*\else#1\NAT@spacechar\fi}%
       {\@citeb\@extra@b@citeb}%
     \NAT@date}}
  {\@citea\NAT@nmfmt{\NAT@nm}%
   \NAT@spacechar\NAT@@open\if*#1*\else#1\NAT@spacechar\fi\NAT@hyper@{\NAT@date}}
  {}{}
\makeatother

\newcommand{\JWST}{\textit{JWST}}
\newcommand{\HST}{\textit{HST}}

\newcommand{\hg}{H$\gamma$}
\newcommand{\hb}{H$\beta$}
\newcommand{\ha}{H$\alpha$}
\newcommand{\oiii}{[O\,\textsc{iii}]}
\newcommand{\um}{$\mu$m}

\newcommand{\Lya}{Ly$\alpha$}

\newcommand{\civ}{C\,\textsc{iv}}
\newcommand{\heii}{He\,\textsc{ii}}
\newcommand{\ciii}{C\,\textsc{iii}]}
\newcommand{\mgii}{Mg\,\textsc{ii}}

\newcommand{\civA}{C\,\textsc{iv}\,$\lambda\lambda$1549,1551}
\newcommand{\heiiAuv}{He\,\textsc{ii}\,$\lambda 1640$}
\newcommand{\ciiiA}{C\,\textsc{iii}]\,$\lambda\lambda$1907,1909}
\newcommand{\mgiiA}{Mg\,\textsc{ii}\,$\lambda\lambda$2797,2803}

\newcommand{\heiiAopt}{He\,\textsc{ii}\,$\lambda 4687$}
\newcommand{\heiA}{He\,\textsc{i}\,$\lambda 5876$}

\newcommand{\fex}{[Fe\,\textsc{x}]}
\newcommand{\fexA}{[Fe\,\textsc{x}]\,$\lambda 6376$}
\newcommand{\siii}{Si\,\textsc{ii}}

\newcommand\target{COS-66964}

\begin{document}

\title{\large Strong rest-UV emission lines in a ``little red dot'' AGN at $\boldsymbol{z=7}$: \\ Early SMBH growth alongside compact massive star formation?}

\shortauthors{Akins et al.}
\shorttitle{}

\correspondingauthor{Hollis B. Akins} 
\email{hollis.akins@gmail.com}

\author[0000-0003-3596-8794]{Hollis B. Akins}
\altaffiliation{NSF Graduate Research Fellow}
\affiliation{Department of Astronomy, The University of Texas at Austin, 2515 Speedway Blvd Stop C1400, Austin, TX 78712, USA}

\author[0000-0002-0930-6466]{Caitlin M. Casey}
\affiliation{Department of Astronomy, The University of Texas at Austin, 2515 Speedway Blvd Stop C1400, Austin, TX 78712, USA}
\affiliation{Cosmic Dawn Center (DAWN), Denmark}

\author[0000-0002-4153-053X]{Danielle A. Berg}
\affiliation{Department of Astronomy, The University of Texas at Austin, 2515 Speedway Blvd Stop C1400, Austin, TX 78712, USA}

\author[0000-0002-0302-2577]{John Chisholm}
\affiliation{Department of Astronomy, The University of Texas at Austin, 2515 Speedway Blvd Stop C1400, Austin, TX 78712, USA}

\author[0000-0002-3560-8599]{Maximilien Franco} 
\affiliation{Department of Astronomy, The University of Texas at Austin, 2515 Speedway Blvd Stop C1400, Austin, TX 78712, USA}

\author[0000-0001-8519-1130]{Steven L. Finkelstein} 
\affiliation{Department of Astronomy, The University of Texas at Austin, 2515 Speedway Blvd Stop C1400, Austin, TX 78712, USA}

\author[0000-0002-7530-8857]{Seiji Fujimoto}
\altaffiliation{Hubble Fellow}
\affiliation{Department of Astronomy, The University of Texas at Austin, 2515 Speedway Blvd Stop C1400, Austin, TX 78712, USA}

\author[0000-0002-5588-9156]{Vasily Kokorev}
\affiliation{Department of Astronomy, The University of Texas at Austin, 2515 Speedway Blvd Stop C1400, Austin, TX 78712, USA}

\author[0000-0003-3216-7190]{Erini Lambrides}
\affiliation{NASA Goddard Space Flight Center, 8800 Greenbelt Rd, Greenbelt, MD 20771, USA}
\altaffiliation{NPP Fellow}

\author[0000-0002-4271-0364]{Brant E. Robertson}
\affiliation{Department of Astronomy and Astrophysics, University of California, Santa Cruz, 1156 High Street, Santa Cruz, CA 95064, USA}

\author[0000-0003-1282-7454]{Anthony J. Taylor}
\affiliation{Department of Astronomy, The University of Texas at Austin, 2515 Speedway Blvd Stop C1400, Austin, TX 78712, USA}

\author[0000-0003-4263-2228]{David~A.~Coulter}
\affiliation{Space Telescope Science Institute, Baltimore, MD 21218, USA}


\author[0000-0002-4924-444X]{Ori Fox}
\affiliation{Space Telescope Science Institute, Baltimore, MD 21218, USA}

\author[0000-0003-2495-8670]{Mitchell Karmen}
\altaffiliation{NSF Graduate Research Fellow}
\affiliation{Physics and Astronomy Department, Johns Hopkins University, Baltimore, MD 21218, USA}

%




\begin{abstract} 
\textit{JWST} has now revealed a population of broad-line AGN at $z\gtrsim4$ characterized by a distinctive SED shape, with very red rest-frame optical and very blue rest-frame UV continuum. While the optical continuum is thought to originate from the accretion disk, the origin of the UV continuum has been largely unclear. We report the detection of the strong rest-frame UV emission lines of C\,\textsc{iii}]\,$\lambda\lambda$1907,1909 and C\,\textsc{iv}\,$\lambda\lambda$1549,1551 in a ``little red dot'' AGN, COS-66964. Spectroscopically confirmed at $z=7.0371$, COS-66964 exhibits broad H$\alpha$ emission (FWHM $\sim 2000$\,km\,s$^{-1}$), and weak broad H$\beta$, implying significant dust attenuation to the BLR ($A_V = 3.9^{+1.7}_{-0.9}$). The H$\alpha$ line width implies a central SMBH mass of $M_{\rm BH} = \left(1.9^{+1.6}_{-0.7}\right)\times10^{7}$\,M$_\odot$, and an Eddington ratio $\lambda\sim0.3$--$0.5$. While marginal He\,\textsc{ii}\,$\lambda4687$ and [Fe\,\textsc{x}]\,$\lambda6376$ detections further indicate that the AGN dominates in the rest-frame optical, the non-detection of He\,\textsc{ii}\,$\lambda1640$ in the UV despite high EW C\,\textsc{iii}] and C\,\textsc{iv} ($\sim 35$\,\AA) is more consistent with photoionization by massive stars. The non-detection of Mg\,\textsc{ii}\,$\lambda\lambda$2800 is similarly inconsistent with an AGN scattered light interpretation. Assuming the rest-frame UV is dominated by stellar light, we derive a stellar mass of $\log M_\star/M_\odot\sim8.5$, implying an elevated $M_{\rm BH}/M_\star$ ratio $\sim2$ orders of magnitude above the local relation, but consistent with other high-$z$ AGN discovered by \textit{JWST}. The source is unresolved in all bands, implying a very compact size $\lesssim200$ pc in the UV. This suggests that the simultaneous buildup of compact stellar populations (i.e., galaxy bulges) and the central SMBH is ongoing even at $z\gtrsim7$. 
\end{abstract}

\section{Introduction}\label{sec:intro}

The study of the first supermassive black holes (SMBHs) and active galactic nuclei (AGN) is particularly important for our understanding of the assembly of galaxies in the Universe \citep{fabianObservational2012,kormendyCoevolution2013a,heckmanCoevolution2014a}. 
Despite the fact that UV-bright quasars with black hole masses $>10^9$\,M$_\odot$ are already in place at $z>7$ \citep{wangLuminous2021}, it remains unclear what the dominant seed population is \citep[i.e.,~stellar vs. direct-collapse black hole seeds;][]{brommFormation2003, agarwalUbiquitous2012, johnsonSupermassive2013, smithSupermassive2019, inayoshiAssembly2020}, how they grow \citep[i.e., sub vs.~super-Eddington accretion;][]{pezzulliSuperEddington2016, reganSuperEddington2019, massonneauHow2023}, and how they impact their host galaxies and the IGM \citep{fanQuasars2023}.

\JWST\ has already made major advances towards building a more complete picture of AGN at high-redshift. 
Numerous high-$z$ AGN have been identified and confirmed via broad Balmer emission lines \citep{harikaneJWST2023,larsonCEERS2023, ublerGANIFS2023, maiolinoJADES2023, taylorBroadLine2024}, exotic high-ionization lines \citep{scholtzJADES2023, mazzolariNarrow2024, maiolinoSmall2024, chisholmNe2024}, and X-ray emission associated with \JWST\ counterparts \citep{bogdanEvidence2024, gouldingUNCOVER2023}.
A particular class of AGN discovered  by \JWST\ exhibit very red rest-frame optical colors, so-called ``little red dots'' \citep[LRDs;][]{mattheeLittle2024}.  
These objects are characterized by point-like morphology, ubiquitous broad Balmer lines \citep{mattheeLittle2024, greeneUNCOVER2024, kocevskiRise2024}, and red continuum from rest-frame $3000$\,\AA--$1$\,\um, indicating a direct view to the accretion disk/broad-line region (BLR), but with significant foreground dust attenuation.  
They also often exhibit blue colors in the rest-frame UV ($\sim 1000$--$3000$\,\AA), perhaps indicating a composite galaxy+AGN SED \citep[e.g.][]{akinsTwo2023,barroExtremely2024} or AGN light scattered/leaked through the dust screen \citep[e.g.][]{labbeUNCOVER2023}.

The LRDs have raised a number of issues challenging the canonical AGN paradigm. 
In particular, they are remarkably abundant, comprising some $20\%$ of all broad-line AGN observed at $z\gtrsim 4$ \citep{harikaneJWST2023, taylorBroadLine2024}. 
Moreover, when accounting for dust attenuation, the LRDs appear to dominate the bolometric luminosity function for high-$z$ AGN, with volume densities $\sim 10$--$100$ times higher than UV-bright quasars at the same bolometric luminosity \citep{greeneUNCOVER2024, kokorevCensus2024, akinsCOSMOSWeb2024}. 
They also generally exhibit weak near-IR/mid-IR emission, in contrast to what would be expected from canonical hot dust torus models \citep[][Leung et al.~\textit{in prep.}]{williamsGalaxies2023, perez-gonzalezWhat2024, akinsCOSMOSWeb2024}, they are generally not X-ray detected, even in deep stacks \citep[]{anannaXray2024, yueStacking2024, maiolinoJWST2024, lambridesCase2024}, and do not exhibit significant variability, despite their low masses \citep{kokuboChallenging2024}. 
These results suggest that either the LRDs are a unique population of AGN which defy our existing picture of AGN unification, or that the AGN contribution to the LRDs is overestimated, perhaps due to the failure of our empirical calibrations for black hole mass/bolometric luminosity.

A possible solution to the tensions posed by the AGN interpretation of LRDs is the possibility that a significant portion of the emission originates from stars. 
In fact, early \JWST\ studies searching for high-$z$ massive galaxy candidates identified many LRD-like objects as massive, dust-obscured or Balmer break candidates \citep[e.g.][]{labbePopulation2023, akinsTwo2023}. 
With very small effective radii, these objects would represent remarkably dense/compact galaxies \citep[see e.g.][]{baggenSizes2023}. 
Some of these candidates have since been found to indeed show Balmer break features in their spectra, as well as broad Balmer lines \citep{wangRUBIES2024}, implying a significant contribution from both evolved stars and AGN in the rest-frame optical (though, see \citet{inayoshiExtremely2024} for an alternative interpretation in which the Balmer break arises from extremely dense gas near the AGN).

In this letter we present spectroscopic observations of \target, a ``little red dot'' AGN now confirmed at $z=7.0371$.
\target\ was previously reported in \citet{kocevskiRise2024} as PRIMER-COS-7103 with a photometric redshift of 7.03.
In Section~\ref{sec:data} we describe the \JWST/NIRCam and NIRSpec data used. 
In Section~\ref{sec:results} we present the spectrum and derived properties, and in Section~\ref{sec:discussion} we discuss the implications of our results.
Throughout this paper, we adopt a \citet{kroupaInitial2002} initial mass function and a cosmology consistent with the \citet{planckcollaborationPlanck2020} results ($H_0=67.66$ km s$^{-1}$ Mpc$^{-1}$, $\Omega_{m,0} = 0.31$). 
All magnitudes are quoted in the AB system \citep{okeAbsolute1974}.

\section{Data}\label{sec:data}

\subsection{JWST/NIRCam imaging}

\target\ falls within the Public Release Imaging for Extragalactic Research (PRIMER) survey (P.I.~J.~Dunlop, GO\#1837) in the COSMOS field. 
PRIMER is a large Cycle 1 Treasury Program to image two \HST\ CANDELS Legacy Fields (COSMOS and UDS) with NIRCam+MIRI \citep{donnanJWST2024}. 
The PRIMER-COSMOS field comprises $\sim 130$ sq.~arcmin of NIRCam imaging in F090W, F115W, F150W, F200W, F277W, F356W, F410M, and F444W, plus $\sim 110$ sq. arcmin of MIRI imaging in F770W and F1800W. 
The raw NIRCam imaging was reduced by the \JWST\ Calibration Pipeline version 1.12.1, with the addition of several custom modifications \citep[as has also been done for other \JWST\ studies, e.g.][]{bagleyCEERS2022} including the subtraction of $1/f$ noise and sky background. 
We use the Calibration Reference Data System (CRDS)\footnote{\url{https://jwst-crds.stsci.edu/}} pmap 1170 which corresponds to NIRCam instrument mapping imap 0273.
The final mosaics are created in Stage 3 of the pipeline with a pixel size of $0\farcs03$/pixel. 
Astrometric calibration is conducted via the \JWST/\HST alignment tool \citep[JHAT,][]{restJWST2023}, with a reference catalog based on an \HST/F814W $0\farcs03$/pixel mosaic in the COSMOS field \citep{koekemoerCOSMOS2007} with astrometry tied to Gaia-EDR3 \citep{gaiacollaborationGaia2018}. The median offset in RA and Dec between our reference catalog and the NIRCam mosaic is less than 5 mas.
\target\ does not fall within the MIRI coverage in the PRIMER survey, nor in the MIRI/F770W coverage from COSMOS-Web \citep[GO\#1727,][]{caseyCOSMOSWeb2023}, which covers much of the same field.

\subsection{JWST/NIRSpec spectroscopy}

\target\ was observed with \JWST/NIRSpec PRISM spectroscopy as part of director's discretionary time program \#6585 (P.I.~D.~Coulter). 
The program was primarily intended to target high-$z$ supernova candidates, identified via NIRCam difference imaging in the overlapping area of PRIMER and COSMOS-Web. 
\target\ was included among $\sim 300$ high-$z$ galaxy filler targets, which were split across three dithers according to priority. 
The source was included in two of three dithers, for a total exposure time of 12080s (3.35hr) in the PRISM mode.

We reduce the NIRSpec data using the standard \JWST\ pipeline (version 1.14.0), with the addition of improved snowball correction via the \texttt{snowblind} package\footnote{\url{https://github.com/mpi-astronomy/snowblind}}.
The pipeline produces 1D and 2D spectroscopic outputs. 
We manually extract the 1D spectrum from the 2D pipeline output to optimize the detection signal-to-noise. 
We define a custom extraction kernel based on the cross-dispersion profile of the bright \Lya, \oiii, and \ha\ emission lines, from which we extract the 1D spectrum across the full wavelength range following \citet{horneOptimal1986}. 
We note that analysis of NIRSpec MOS data requires careful consideration of slit-losses, as the micro-shutters are often smaller than the size of the targets.
Even for point sources (as is the case here), the NIRSpec PSF can be larger than the micro-shutter size, and targets may not be centered in the shutter, leading to significant slit loss.  
The pipeline includes an automatic slit loss correction, though we disable this step in favor of an empirical calibration to match the observed NIRCam photometry.
By convolving the optimally-extracted PRISM spectrum with the NIRCam filter curves, we find that the necessary slit-loss correction is $\sim 0.8$, consistent across all bands.

\begin{figure*}
\centering
\includegraphics[width=\linewidth]{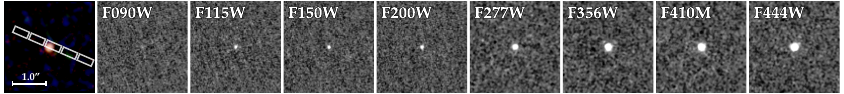}
\includegraphics[width=\linewidth]{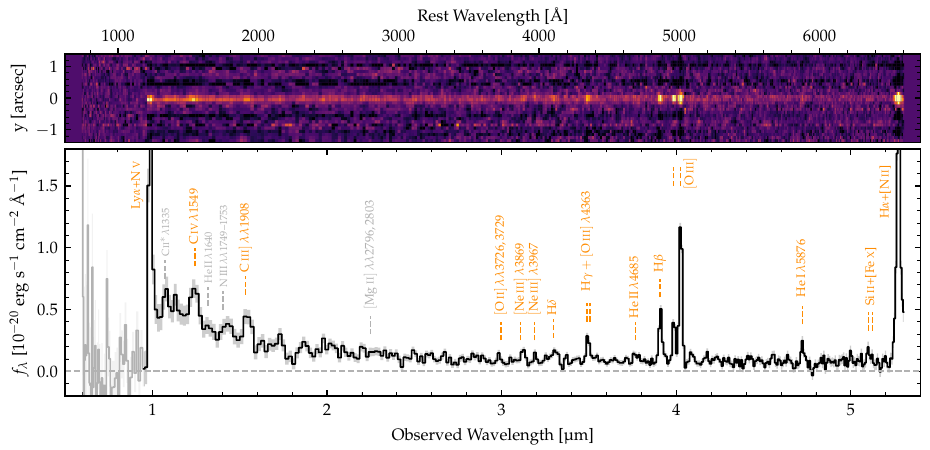}
\caption{\JWST/NIRCam photometry and the NIRSpec PRISM spectrum for \target. \textbf{Top}: Cutouts in the 8 NIRCam bands available from PRIMER, as well as an RGB image. We also highlight the position of the NIRSpec/MSA shutters over the RGB image. 
\textbf{Bottom:} The 2D and optimally-extracted 1D spectrum. The 1$\sigma$ uncertainty on the spectrum is indicated with the grey shaded region. Several notable emission lines are marked, including \Lya, \civ, \ciii, \hb, \oiii, and \ha. Emission lines labeled in grey are not significantly detected (see \S\ref{sec:methods}). The continuum is detected across the full wavelength range, and exhibits a turnover from blue to red around $3$\,\um\ (rest-frame 3800\,\AA).}\label{fig:spectrum}	
\end{figure*}

\section{Results}\label{sec:results}

\subsection{Strong rest-UV lines in a ``little red dot'' at $z=7.0371$}

Figure~\ref{fig:spectrum} shows the 2D and optimally-extracted 1D spectrum for \target. 
The 1$\sigma$ uncertainty on the PRISM spectrum is indicated with the grey shaded region, and we label several notable emission lines, including \Lya, \civ, \ciii, \hb, \oiii, and \ha. 
We additionally show cutouts in the 8 PRIMER NIRCam bands, and an RGB image highlighting the position of the NIRSpec shutters.

Based on the detection of strong \oiii\,$\lambda\lambda4959$,$5007$ and \ha\ emission, we determine a spectroscopic redshift of $z_{\rm spec} = 7.0371^{+0.0006}_{-0.0005}$. 
This places \target\ among the highest redshift confirmed LRDs \citep[see e.g.][]{kokorevUNCOVER2023, greeneUNCOVER2024, furtakHigh2024, wangRUBIES2024}. 
We note that \ha\ is cut off at the red end of the detector, given the redshift of the source, though most of the line is still detected, allowing kinematic decomposition.

Notably, we also detect strong \Lya\ emission (blended with N\,\textsc{v}), \civA, and \ciiiA\ emission. 
\Lya\ has been detected in several LRDs, even at $z\gtrsim 7$, despite the opacity of the IGM at these redshifts \citep[see e.g.][]{kokorevUNCOVER2023,furtakHigh2024}, which likely indicates that these objects reside in ionized bubbles. 
However, the high ionization lines of \civ\ and \ciii\ are not typically observed in LRDs; for the most part, their UV spectra appear featureless \citep[e.g.][]{greeneUNCOVER2024}. 
We also note the presence of weak emission features at the expected locations of \heiiAopt\ and \siii+\fexA\ in the optical.

In addition to the detected emission lines, we note that the continuum is detected across the entire wavelength range from $1$--$5$\,\um. 
The continuum transitions from blue to red around 3\,\um\ (rest-frame $3800$\,\AA). 
We note, however, that we don't observe a clear Balmer break, which would indicate the presence of an old stellar population.

\subsection{Spectral Fitting and Line Decomposition}\label{sec:methods}

We fit the \JWST/NIRSpec PRISM spectrum of \target\ with a custom, flexible galaxy+AGN SED model using the \textsc{UltraNest} nested sampling package \citep{buchnerStatistical2016,buchnerCollaborative2019,buchnerUltraNest2021}.  
In our model, emission lines are handled separately from the continuum, to directly fit for line fluxes and allow more flexibility in the physical models.

For the galaxy (continuum) model, we use the BPASS library of stellar SED models \citep{eldridgeBinary2017} and a non-parametric SFH model \citep{lejaHow2019}. 
The SFH is parametrized by the $\Delta\log({\rm SFR})$ in adjacent time bins; we adopt the ``bursty continuity'' prior \citep[described in][]{tacchellaStellar2022}, i.e. the prior on $\Delta\log({\rm SFR})$ is a $t$-distribution with $\sigma = 1$ and $\nu = 2$ degrees of freedom. 
We adopt four fixed age bins from $0$--$10$, $10$--$50$, $50$--$200$, and $200$--$400$ Myr. 
We adopt log-uniform priors on the stellar mass (from $10^6$ to $10^{13}~M_\odot$) and metallicity (from $10^{-3}$ to $0.5~Z_\odot$).
For the galaxy model, we adopt an SMC dust law with $A_V$ allowed to vary from $10^{-3}$ to $0.3$ (with a log-uniform prior) and include nebular continuum using pre-computed \textsc{cloudy} grids (\citealt{bylerNebular2017}; though we do not model lines with \textsc{cloudy}, as these are handled separately).
We model the AGN continuum as a simple power law with a fixed slope $\beta = -7/3 = -2.33$ and intrinsic (i.e.,~unattenuated) UV magnitude from $M_{\rm UV} \sim -25$ to $-19$. 
We include dust attenuation following the \citet{salimDust2018} model, with a power law index from $\delta$ from $-0.6$ to $+0.2$, with a Gaussian prior centered on $\delta = -0.45\pm 0.1$ (roughly an SMC law) and $A_V$ allowed to vary from $0.5$ to $6.0$. 
Note that our choice of the range on $A_V$ for the galaxy/AGN components restricts the model to a scenario in which the galaxy dominates in the rest-frame UV, while the AGN dominates in the rest-frame optical. 
We motivate this decision based on the emission line ratios in Section~\ref{sec:ratios}, and while we focus primarily on the emission lines in the following sections, we return to discuss the continuum decomposition in Section~\ref{sec:galagn}.

Emission lines are then added on top of the continuum and modeled as Gaussians. 
We include all lines annotated in Figure~\ref{fig:spectrum} and fit simultaneously for their fluxes with their positions and widths tied together. 
The lines are split into two groups---narrow and broad---with a single FWHM fit for each group. 
The narrow line widths are allowed to vary from $100$ to $300$ km/s, while the broad line widths vary from $700$ to $3000$ km/s.
We only include broad components for \hb, \heiA, and \ha. 
We note that we model H$\gamma$ and \oiii\,$\lambda$4363 separately, though the lines are blended. 
we include the [N\,\textsc{ii}]\,$\lambda\lambda6548,6583$ doublet with a fixed line ratio of 1:3, and we fix the ratio of \oiii\,$\lambda$5007/\oiii\,$\lambda$4959 to 3 and [Ne\,\textsc{iii}]\,$\lambda$3967/[Ne\,\textsc{iii}]\,$\lambda$3869 to 0.3.

A number of calibration effects are incorporated directly into our model fit. 
For one, we fit for a slight velocity offset in \ha\ (between $-500$ and $+500$ km/s), as the wavelength calibration at the very red end is not perfect.
Moreover, in each iteration of the model fit, the internal model spectrum is convolved with the line spread function for the NIRSpec/PRISM disperser. 
For a uniformly illuminated slit, the PRISM resolution varies from $R\sim 70$ at the blue end to $R\sim 300$ at the red end. 
To account for this, the model spectrum is computed on a wavelength grid sampled uniformly in $1/R$ space (based on the published \texttt{JDox} PRISM resolution curve). 
The spectrum is then convolved with a constant Gaussian kernel and interpolated to the wavelength grid of the PRISM data. 
As noted in \citet{degraaffIonised2024}, the normalization of the resolution curve for NIRSpec/MOS data depends strongly on the source morphology; for a point source, it can be up to twice the reported resolution. 
We therefore implement a nuisance parameter $f_{\rm LSF}$ which scales the line spread function (i.e., the width of the Gaussian convolution kernel); we find a best-fit value $\sim 1.3$, i.e. the maximum resolution is $\sim 400$, consistent with findings in \citet{furtakHigh2024,degraaffIonised2024}. 
This parameter is constrained primarily by the $[{\rm O}\,\textsc{iii}]\,\lambda\lambda 4959,5007$ doublet, which is resolved in our data (Fig.~\ref{fig:spectrum}), but wouldn't be at the nominal PRISM resolution.

\begin{deluxetable}{@{\extracolsep{5pt}}l@{}C@{}C@{}C@{}}
\tabletypesize{\small}
\centering
\tablecaption{Measured line fluxes and EWs.}\label{tab:fluxes}
\tablehead{\colhead{Line} & \colhead{$\lambda_{\rm rest}$} & \colhead{Flux$\,\times\,10^{20}$} & \colhead{EW$_{\rm rest}$} \\[-0.5em] 
\colhead{(component)} & \colhead{[\AA]} & \colhead{[erg\,s$^{-1}$\,cm$^{-2}$]} & \colhead{[\AA]}}
\startdata
Ly$\alpha$ & 1215.670 & 57.2^{+3.1}_{-3.4} &265^{+37}_{-31} \\
${\rm C}\,\textsc{ii}^*\lambda1335$ & 1335.708 & < 5.1 &< 11 \\
${\rm C}\,\textsc{iv}\,\lambda\lambda1549,1551$ & 1549.480 & 13.8^{+2.5}_{-2.8} &34^{+6}_{-7} \\
He\,\textsc{ii}\,$\lambda1640$ & 1640.400 & < 3.8 &< 10 \\
N\,\textsc{iii}\,$\lambda\lambda 1749$--$1753$ & 1749.246 & < 3.9 &< 12 \\
${\rm C}\,\textsc{iii}]\,\lambda\lambda1908$ & 1908.734 & 9.2^{+2.3}_{-2.0} &34^{+9}_{-8} \\
Mg\,\textsc{ii}\,$\lambda\lambda2797,2803$ & 2799.942 & < 1.7 &< 13 \\
$[{\rm O}\,\textsc{ii}]\,\lambda\lambda 3726,3729$ & 3728.484 & 0.9^{+0.5}_{-0.5} &11^{+6}_{-6} \\
$[{\rm Ne}\,\textsc{iii}]\,\lambda 3869$ & 3869.857 & 1.4^{+0.6}_{-0.6} &15^{+6}_{-6} \\
$[{\rm Ne}\,\textsc{iii}]\,\lambda 3967^{*}$ & 3968.593 & 0.4^{+0.2}_{-0.2} &5^{+2}_{-2} \\
${\rm H}\epsilon$ & 3971.202 & 1.1^{+0.5}_{-0.6} &12^{+6}_{-6} \\
${\rm H}\delta$ & 4102.900 & 2.4^{+0.6}_{-0.5} &28^{+7}_{-6} \\
${\rm H}\gamma$ & 4341.692 & 4.6^{+0.6}_{-0.6} &56^{+8}_{-7} \\
$[{\rm O}\,\textsc{iii}]\,\lambda 4363$ & 4364.437 & 2.3^{+0.5}_{-0.6} &27^{+7}_{-7} \\
He\,\textsc{ii}\,$\lambda4687$ & 4687.022 & 1.7^{+0.6}_{-0.5} &20^{+7}_{-6} \\
${\rm H}\beta$ (narrow) & 4862.692 & 8.8^{+1.3}_{-1.2} &116^{+19}_{-16} \\
${\rm H}\beta$ (broad) & 4862.692 & 3.4^{+1.5}_{-1.7} &45^{+20}_{-22} \\
$[{\rm O}\,\textsc{iii}]\,\lambda 4959$ & 4960.296 & 9.9^{+0.2}_{-0.2} &123^{+5}_{-4} \\
$[{\rm O}\,\textsc{iii}]\,\lambda 5007^\dagger$ & 5008.241 & 29.6^{+0.7}_{-0.7} &372^{+15}_{-13} \\
He\,\textsc{i}\,$\lambda5876$ (narrow) & 5877.255 & 2.9^{+0.7}_{-0.7} &39^{+10}_{-9} \\
He\,\textsc{i}\,$\lambda5876$ (broad) & 5877.255 & < 1.7 &< 23 \\
${\rm Si}\,\textsc{ii}$ & 6348.858 & 2.3^{+0.6}_{-0.7} &31^{+9}_{-9} \\
$[{\rm Fe}\,\textsc{x}]$ & 6376.275 & \left(1.0^{+0.6}_{-0.5} \right) & \left(14^{+9}_{-8} \right) \\
$[{\rm N}\,\textsc{ii}]\,\lambda 6549$ & 6549.862 & < 0.6 &< 8 \\
${\rm H}\alpha$ (narrow) & 6564.635 & 31.5^{+2.7}_{-2.4} &446^{+49}_{-45} \\
${\rm H}\alpha$ (broad) & 6564.635 & 51.2^{+3.1}_{-3.6} &725^{+72}_{-72} \\
$[{\rm N}\,\textsc{ii}]\,\lambda 6585^\ddagger$ & 6585.282 & < 1.8 &< 24 \\
\enddata
\tablenotetext{*}{[Ne\,\textsc{iii}]\,$\lambda 3967$$/$[Ne\,\textsc{iii}]\,$\lambda 3869$ is fixed to 0.3.\vspace{-5pt}}
\tablenotetext{\dagger}{[O\,\textsc{iii}]\,$\lambda 5007$$/$[O\,\textsc{iii}]\,$\lambda 4959$ is fixed to 3.0.\vspace{-5pt}}
\tablenotetext{\ddagger}{[N\,\textsc{ii}]\,$\lambda 6585/$[N\,\textsc{ii}]\,$\lambda 6549$ is fixed to 3.0. \vspace{-5pt}}
\end{deluxetable}
\vspace{-12pt}

\begin{figure*}
\centering
\includegraphics[width=\linewidth]{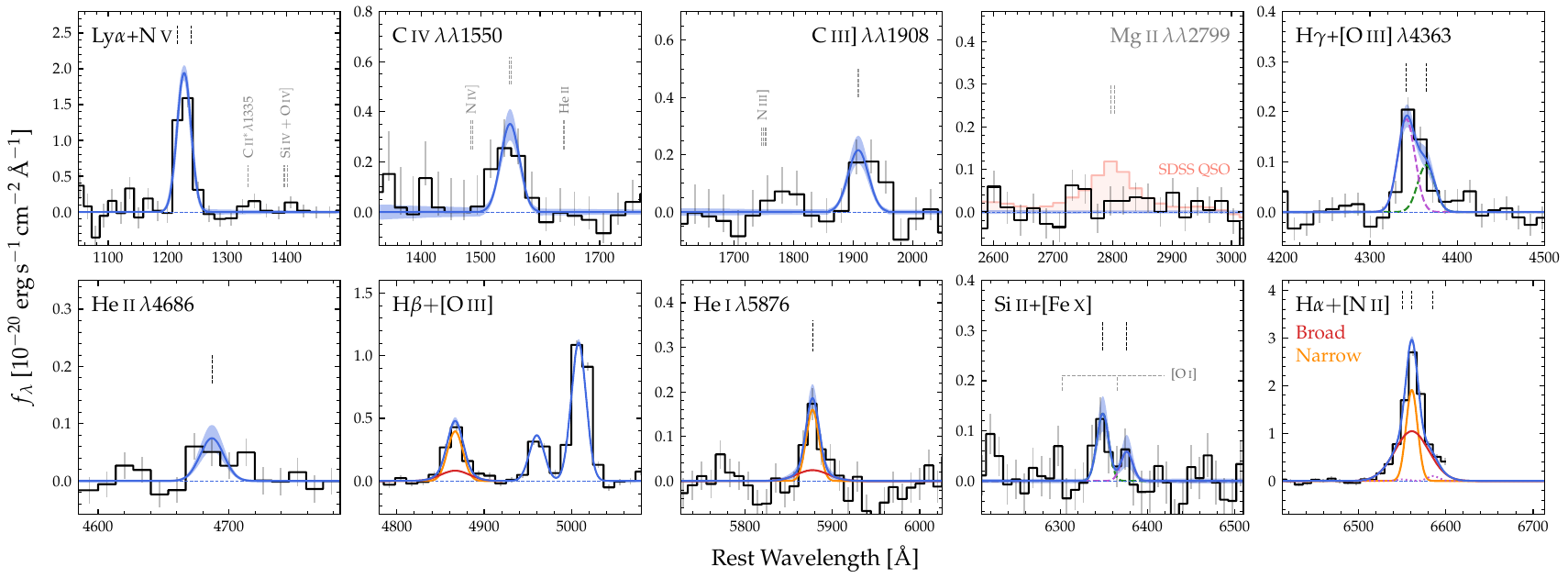}
\caption{Zoom-in around several emission lines of interest in the NIRSpec/PRISM spectrum of \target. In all panels, the spectrum has been continuum-subtracted using the best-fit model.
Blue lines and shaded regions show the posterior model spectrum, and include the uncertainty on the subtracted continuum. 
For \hb, \heiA, and \ha, we show the narrow+broad line decomposition in orange and red, respecitvely. We derive narrow/broad FWHMs of $200^{+50}_{-50}.$ and $2010^{+130}_{-120}$ km\,s$^{-1}$, respectively. The blended lines \hg+\oiii\,$\lambda$4363 and \siii+\fex\ are fit with multiple Gaussians, which are plotted separately in purple/green dashed lines. For \mgii, we overplot the SDSS QSO composite \citep{vandenberkComposite2001} scaled to match the observed continuum and convolved to the PRISM resolution. }\label{fig:emissionlines}	
\end{figure*}

Figure~\ref{fig:emissionlines} zooms in on various emission lines of interest in the spectrum. 
Here, we show the results of our line fitting in blue; shaded regions indicate the 1$\sigma$ confidence on the posterior spectrum. 
We confirm the significance of the detections of \civ\ and \ciii\ (${\rm S}/{\rm N} \sim 3.8$--$4.4$). 
Notably, we do not detect \heiiAuv, a line commonly observed in AGN, nor the high ionization line N\textsc{iv}]\,$\lambda 1490$.
We also do not detect \mgiiA\ emission, which is often observed in Type I AGN \citep[e.g.][EW $\sim 30$\,\AA, shown in Fig.~2]{vandenberkComposite2001} and commonly used as a virial tracer of the black hole mass \citep[e.g.][]{wangEstimating2009}. 
We place a limit on the rest-frame equivalent width of \mgii\ of $<13$\,\AA\ (fluxes and EWs of all lines are given in Table~\ref{tab:fluxes}).
The non-detections of these lines appear at odds with the AGN interpretation; we return to this question shortly.

We also show in Figure~\ref{fig:emissionlines} the broad+narrow line decomposition for \hb, \heiA, and \ha. 
While \ha\ falls at the very red end of the spectrum, and is cut off at $5.3$\,\um, we see a very clear broad component, with a width of FWHM $\sim 2010^{+130}_{-120}$ km\,s$^{-1}$. 
The broad component is robustly detected: without it, the fit is significantly worse, with a difference in the Bayesian Information Criterion (BIC) of $\Delta$BIC $\gtrsim 100$. 
The broad component in H$\beta$ is less significant, and we detect no broad component in \heiA, though the line is much weaker. 
We note that we also do not detect broad components in \oiii\ ($\Delta$BIC $\sim 15$). 
We assume that the weak broad H$\beta$ is due to significant dust attenuation towards the BLR, though we discuss alternative scenarios in \S\ref{sec:discussion}.  
Adopting an intrinsic broad H$\alpha$/H$\beta$ flux ratio of 3.06 \citep{dongBroadline2007} and assuming an SMC-like extinction curve, we derive an attenuation $A_V \approx 3.9^{+1.7}_{-0.9}$. 
Similarly, for the narrow line region (assuming an intrinsic ratio of 2.86 consistent with case-B recombination) we derive $A_V \approx 0.4^{+0.4}_{-0.4}$.

From the dust-corrected broad H$\alpha$ luminosity, we compute the black hole mass following the standard single-epoch virial black hole mass calibration from \citet{greeneEstimating2005}. 
We derive a black hole mass of $1.9^{+1.6}_{-0.7}\times 10^{7}$ M$_\odot$. 
We additionally derive a bolometric luminosity from \ha\ of $\log L_{{\rm bol,H}\alpha}/{\rm erg}\,{\rm s}^{-1} = 44.8^{+0.6}_{-0.5}$ assuming a bolometric correction of $130\pm 2.4$ \citep{sternType2012}. 
The Eddington ratio is therefore estimated to be $\lambda_{{\rm Edd,H}\alpha} = 0.3^{+0.5}_{-0.2}$, below the Eddington limit.

Finally, we highlight the high-ionization lines of \heiiAopt\ and \fexA\ in the optical. 
While \heiiAopt\ is detected at ${\rm S}/{\rm N} \sim 3$, \fex\ is only marginally detected, with ${\rm S}/{\rm N} \sim 2$, and is somewhat blended with \siii\,$\lambda$6349. \fex\ may also be contaminated by [O\,\textsc{i}]\,$\lambda$6365, though the profile is better fit by \fex+\siii\ alone.

\begin{figure*}
\centering
\includegraphics[width=\linewidth]{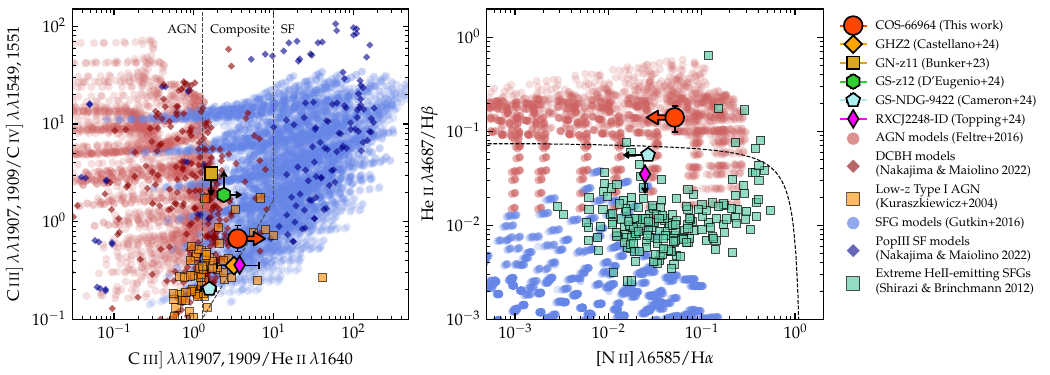}
\caption{Line ratio diagnostics in the UV and optical. In both panels, model grids for AGN and star-forming galaxies from \citet{feltreNuclear2016} and \citet{gutkinModelling2016} are shown in red and blue, respectively. \textbf{Left:} The \ciii/\civ\ vs.~\ciii/\heii\ UV line ratio diagnostic diagram. We plot the classification regions from \citet{scholtzJADES2023} with dashed lines, and we additionally include PopIII and DCBH models from \citet{nakajimaDiagnostics2022} and observed low-$z$ Type I AGN from \citet{kuraszkiewiczEmission2004}.
\target\ is shown in red, where we adopt the 95th percentile of the \heii\ flux as a 2$\sigma$ upper limit. 
Several high-redshift sources are also shown, including GN-z11 \citep{bunkerJADES2023}, GHZ2 \citep{castellanoJWST2024}, GS-z12 \citep{deugenioJADES2023}, RXCJ2248-ID \citep{toppingMetalpoor2024}, and GS-NDG-9422 \citep{cameronNebular2024}. 
Based on the non-detection of \heiiAuv, the UV emission from \target\ lies in the composite region and is consistent with ionization from massive/low-metallicity stars. \textbf{Right:} The \heiiAopt/\hb\ vs.~[N\,\textsc{ii}]/\ha\ diagram. Model grids are the same as in the left panel, and we additionally plot extreme \heii-emitting SFGs from SDSS \citep{shiraziStrongly2012}. Based on the \heiiAopt/\hb\ ratio, the rest-optical emission from \target\ is inconsistent with star formation---even the the most extreme \heii-emitting WR galaxies---and is better fit by ionization from the AGN.}\label{fig:ratios}
\end{figure*}

\subsection{Rest-UV and optical emission line ratios}\label{sec:ratios}

\target\ is one of the only LRDs with significantly detected rest-frame UV emission lines. 
But what do the lines tell us about the nature of the LRDs?

Figure~\ref{fig:ratios} shows two line ratio diagnostic diagrams in the UV and optical. 
First, the left panel shows the \ciii/\civ\ vs.~\ciii/\heiiAuv\ diagram, which has been proposed as a discriminator between AGN and star-formation-powered photoionization. 
In particular, the \heiiAuv\ and \civA\ lines, with ionization potentials 54.4 eV and 47.9 eV, probe the shape of the ionizing continuum. 
We plot AGN and SFG model grids from \citet{feltreNuclear2016} and \citet{gutkinModelling2016}, as well as the demarcation lines for AGN/SF from \citet{scholtzJADES2023}. 
We also plot several notable objects with well characterized UV spectra from \JWST; GN-z11 at \citep[$z=10.6$,][]{bunkerJADES2023}, GHZ2 \citep[$z=12.3$,][]{castellanoJWST2024}, GS-z12 \citep[$z=12.5$,][]{deugenioJADES2023}, RXCJ2248-ID \citep[$z=6.1$,][]{toppingMetalpoor2024}, and GS-NDG-9422 \citep[$z=5.9$,][]{cameronNebular2024}. 
These objects are all consistent with star-formation, save GN-z11 which is more consistent with AGN given other high ionization lines and a very high inferred density \citep{maiolinoSmall2024}.

Adopting the 95th posterior percentile as a $2\sigma$ upper limit on the \heii\ flux, we derive \ciii/\heii $\gtrsim 3.5$, which falls in the composite region. 
The lower limit on \ciii/\heii\ is more consistent with photoionization by star-formation rather than the AGN. 
Nevertheless, we cannot completely rule out AGN photoionization, as the limit is also consistent with some low-z Type I AGN with weak \heiiAuv.
The UV spectrum of \target\ appears similar to high-$z$ \civ\ emitters such as RXCJ2248-ID or GHZ2, which are characterized by intense star-formation in a very dense and low-metallicity environment, driving the high ionization parameter. 
Note that we do not include \oiii$\,\lambda\lambda1663$ in our modeling, as it is blended with \heii. 
However, including \oiii\ only serves to lower the upper limit on the \heii\ flux, increasing \ciii/\heii\ and placing \target\ more firmly in the star-forming region.

The right panel of Figure~\ref{fig:ratios} shows the \heiiAopt/\hb\ vs.~[N\,\textsc{ii}]/\ha\ optical line ratio diagnostic diagram. 
Though the [N\,\textsc{ii}]/\ha\ is not directly constrained by our data, we adopt as an upper limit the 95th percentile [N\,\textsc{ii}] returned by our model fit, which includes [N\,\textsc{ii}] and broad \ha. 
We show the same model grids as for the UV, and additionally overplot observed line ratios from extreme \heii-emitting galaxies from SDSS \citep{shiraziStrongly2012}. 
In stark contrast to the UV, the rest-optical emission from \target\ is inconsistent with star formation---even the the most extreme \heii-emitting Wolf-Rayet (WR) galaxies---and is better fit by ionization from the AGN.

\subsection{Galaxy+AGN SED decomposition}\label{sec:galagn}

\begin{figure*}[t!]
\centering
\includegraphics[width=0.9\linewidth]{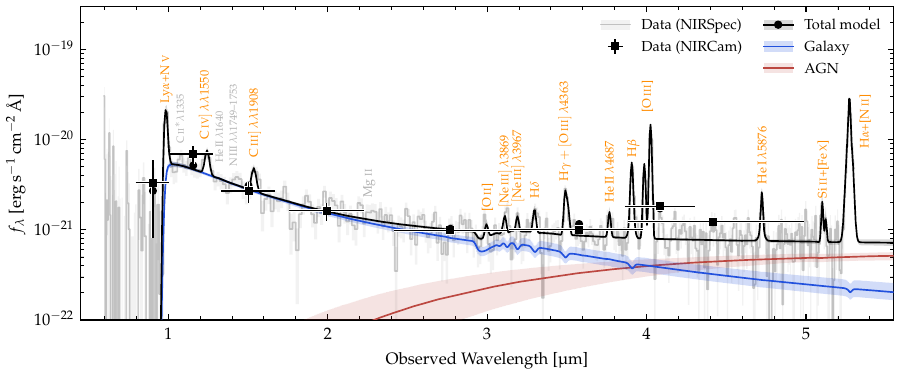}	
\caption{Results from multi-component SED fitting to \target. The NIRSpec/PRISM spectrum is shown in light grey in the background. We show the stellar+nebular continuum model in blue, which dominates in the rest-frame UV, and the AGN continuum model in red, which dominates in the optical. In both cases we show the posterior 16th-84th percentiles with the shaded region. The total model, which includes both continuum components and the fitted nebular lines, is shown in black and matches the observed spectrum and photometry well.  }\label{fig:sedfit}
\end{figure*}

Finally, we return to the continuum decomposition from our full SED model. 
Motivated by the line ratio analysis, which suggests that the UV emission may be dominated by star-formation, while the optical is AGN-dominated, we fit the full SED to a two-component model combining an unobscured galaxy with an obscured AGN continuum (as already described in \S\ref{sec:methods}). 
Figure~\ref{fig:sedfit} shows the resulting SED decomposition; the galaxy component is shown in blue, while the AGN is shown in red. 
We derive a stellar mass of $M_\star = 3.1^{+1.6}_{-1.0} \times 10^8$\,M$_\odot$ with minimal dust attenuation ($A_V \sim 0.1$), consistent with the blue UV slope $\beta_{\rm UV} = -2.1^{+0.2}_{-0.1}$. 
We derive a star-formation rate of ${\rm SFR}_{100} = 1.1^{+0.3}_{-0.3}$ M$_\odot$\,yr$^{-1}$ and a corresponding specific star-formation rate of $\log{\rm sSFR}/{\rm yr}^{-1} = -8.5^{+0.2}_{-0.1}$, consistent with the extrapolation of the star-forming main sequence to $z=7$ \citep{iyerSFRM2018}. 
For the AGN, we derive a continuum bolometric luminosity of $\log L_{\rm bol,cont}/{\rm erg}\,{\rm s}^{-1} = 45.1^{+0.2}_{-0.1}$ assuming a bolometric correction of $5.15$ from $L_{3000}$ \citep{richardsSpectral2006}, consistent with the H$\alpha$ bolometric luminosity. 
The corresponding Eddington ratio is $\lambda_{\rm Edd,cont} = 0.5^{+0.4}_{-0.3}$.

\vspace{-0.35cm}
\begin{deluxetable}{@{\extracolsep{5pt}}l@{}c@{}C@{}}
\tabletypesize{\small}
\centering
\tablecaption{Spectroscopic measurements for \target.}\label{tab:spec}
\tablehead{\colhead{Property} & \colhead{Units} & \colhead{Value}}
\startdata
$z_{\rm spec}$ & \dots & 7.0371^{+0.0006}_{-0.0005} \\
R.A. & hms & 10\mathrm{:}00\mathrm{:}30.1978 \\
Decl. & dms & +02\mathrm{:}23\mathrm{:}22.961 \\
$f_{\lambda,5100}$ & $10^{-22}$\,erg\,s$^{-1}$\,cm$^{-2}$\,\AA$^{-1}$ & 7.9^{+0.2}_{-0.2} \\
$M_{\rm UV}$ & AB mag & -19.17^{+0.05}_{-0.04} \\
$\beta_{\rm UV}$ & \dots & -2.1^{+0.2}_{-0.1} \\
FWHM$_{\rm narrow}$ & km\,s$^{-1}$ & 200^{+50}_{-50} \\
FWHM$_{\rm broad}$ & km\,s$^{-1}$ & 2010^{+130}_{-120} \\
\hline
$A_{V,{\rm broad,H}\alpha/{\rm H}\beta}$ & AB mag & 3.9^{+1.7}_{-0.9} \\
$A_{V,{\rm narrow,H}\alpha/{\rm H}\beta}$ & AB mag & 0.4^{+0.4}_{-0.4} \\
$M_{\rm BH}$ & $10^7$\,M$_{\odot}$ & 1.9^{+1.6}_{-0.7} \\
$\log L_{\rm bol,cont}$ & erg\,s$^{-1}$ & 45.1^{+0.2}_{-0.1} \\
$\log L_{{\rm bol,H}\alpha}$ & erg\,s$^{-1}$ & 44.8^{+0.6}_{-0.5} \\
$\lambda_{\rm Edd,cont}$ & \dots & 0.5^{+0.4}_{-0.3} \\
$\lambda_{{\rm Edd,H}\alpha}$ & \dots & 0.3^{+0.5}_{-0.2} \\
$A_{V,{\rm SED,AGN}}$ & AB mag & 2.4^{+0.6}_{-0.4} \\
\hline
$M_{\star}$ & $10^8$\,M$_\odot$ & 3.1^{+1.6}_{-1.0} \\
SFR$_{100}$ & M$_\odot$ yr$^{-1}$ & 1.1^{+0.3}_{-0.3} \\
$\log$ sSFR$_{100}$ & \dots & -8.5^{+0.2}_{-0.1} \\
$A_{V,{\rm SED,galaxy}}$ & AB mag & 0.12^{+0.04}_{-0.05} \\
\enddata
\end{deluxetable}

We note that \target\ is unresolved in all NIRCam bands. 
We fit the morphology with a simple point source model as well as S\'ersic+point source model in each band to evaluate the significance of any marginally resolved component. 
Forward modeling of the images is performed using \textsc{galsim} \citep{roweGalSim2015} and fitting is performed using the \textsc{MultiNest} nested sampling package \citep{ferozMultimodal2008, ferozMULTINEST2009, ferozImportance2019}, as described in \citet{akinsCOSMOSWeb2024}. 
In all cases, we find that the single point source model is preferred over the  S\'ersic+point source model, though the typical difference in the Bayesian information criterion (BIC) is $\Delta{\rm BIC} \sim 10$, indicating that neither model is a significant improvement over the other.  
This suggests that even though \target\ may be SF-dominated in the rest-UV, the stellar component is very compact, $R_{\rm eff} \lesssim 200$ pc.

\section{Discussion \& Conclusions}\label{sec:discussion}

\begin{figure*}
\centering
\includegraphics[width=\linewidth]{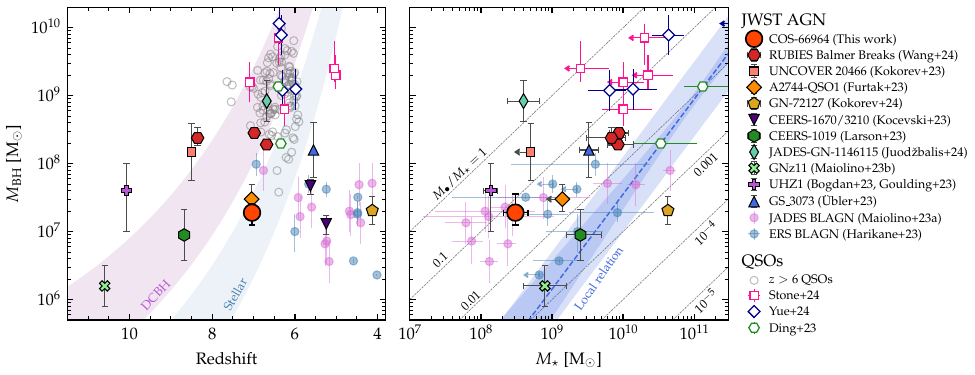}	
\caption{\target\ in context among high-$z$ AGN. \textbf{Left}: Black hole mass vs.~redshift. \target\ is shown in red. Filled points indicate \JWST-selected AGN \citep{wangRUBIES2024,kokorevUNCOVER2023,furtakHigh2024,kokorevSilencing2024,kocevskiHidden2023,larsonCEERS2023,juodzbalisDormant2024,maiolinoSmall2024,bogdanEvidence2024,gouldingUNCOVER2023,maiolinoJADES2023,harikaneJWST2023}. Open points indicate classical QSOs selected from ground-based surveys, from the complilation of \citet[][grey]{fanQuasars2023} and objects with individual host galaxy measurements via 2D decomposition \citep{stoneUndermassive2024, yueEIGER2024, dingDetection2023}. 
We show the range of evolutionary pathways for AGN starting from stellar mass seeds (blue, $M\sim 10$--$300$\,M$_\odot$, $z \sim 15$--$30$) and DCBH seeds (purple, $M\sim 10^{3-5}$\,M$_\odot$, $z \sim 10$--$20$) and accreting at the Eddington limit. \textbf{Right:} Black hole mass vs.~stellar mass. Points are the same as in the left panel. The blue line and shaded region shows the local relation from \citet{reinesRelations2015}. Based on our SED decomposition, \target\ has a $M_{\rm BH}/M_\star$ ratio $\sim 0.1$, consistent with other \JWST-selected AGN at $z\gtrsim 4$.
}\label{fig:mbhz}
\end{figure*}

We have presented the \JWST/NIRSpec PRISM observations of \target, a ``little red dot'' now confirmed at $z_{\rm spec} = 7.0371$. 
We have confirmed the AGN nature of this source via the detection of broad \ha\ (FWHM $\sim 2000$ km\,s$^{-1}$), implying a black hole mass of $M_{\rm BH} \sim 2\times 10^7$ M$_\odot$.
\target\ is unique in its high EW \civA\ emission ($\sim 35$\,\AA), the strongest \civ\ emission in any LRD.
The \civ\ line is often observed in UV-bright quasars \citep[e.g.][]{vandenberkComposite2001}, but rarely seen in star-forming galaxies, typically only being found in dwarf galaxies with very low metallicity and extreme star formation \citep[e.g.][]{starkSpectroscopic2015a,bergIntense2019,toppingMetalpoor2024,izotovExtremely2024}.
Nevertheless, we have shown that the rest-UV line ratios are more consistent with photoionization from intense star-formation, rather than the AGN. 
This is largely due to the non-detection of \heiiAuv, which has a higher ionization potential than \civ\ ($54.4$ vs.~$47.9$ eV). 
This interpretation is supported by the non-detection of \mgiiA, with EW $<13$\,\AA.

By contrast, the broad \ha\ and marginal \heiiAopt\ and \fexA\ detections clearly indicate that the photoionization in the rest-optical is dominated by the AGN. 
\heiiAopt\ is often observed in lower redshift AGN \citep[e.g.][]{kuraszkiewiczEmission2004}, and has been used to trace type II AGN at high-$z$ \citep[e.g.][]{scholtzJADES2023}. 
The detection of \heii\ in the optical, but not the UV, is a clear indicator that the different SED components originate from different processes (in fact, \heiiAopt\ is the weaker of the two lines, with an intrinsic ratio of \heiiAuv/\heiiAopt\ $\sim 7$--$8$ for case-B recombination). 
The \fexA\ coronal line, with an ionization potential of 262 eV, is an even more clear indicator of extremely high ionization gas, which can only be powered by AGN activity, and has also been detected in some LRDs \citep{kocevskiHidden2023, furtakHigh2024}. 
Though both detections are marginal (${\rm S}/{\rm N} \sim 2$--$3$) they are consistent with photoionization from the AGN, at least in the rest-frame optical.

The black hole mass and AGN bolometric luminosity measurements in LRDs generally face significant uncertainty given the unknown nature of these sources. 
We note in particular that the weak broad H$\beta$ could be due to an intrinsically softer ionizing spectrum \citep[perhaps associated with super-Eddington accretion, e.g.][]{pacucciMildly2024,lambridesCase2024}, rather than dust attenuation.
However, the detection of the high-ionization lines \heii\ and \fex\ requires substantial ionizing photon production, inconsistent with the super-Eddington, radiatively inefficient scenario.
This supports the interpretation of the broad \ha/\hb\ ratio as due to dust attenuation.  
The stark difference in $A_V$ for the BLR and NLR ($\sim 4$ vs.~$\sim 0.5$) has been observed in other LRDs \citep[e.g.][]{killiDeciphering2023}, and is consistent with high dust column densities on $\sim 10$ pc scales, perhaps from an extended/dynamic dusty medium in place of a traditional ``torus'' \citep{liLittle2024}. 
We do note, however, that the intrinsic broad H$\alpha$/H$\beta$ ratio is very uncertain due to self-absorption of BLR gas \citep{koristaWhat2004}; the continuum-derived $A_V\sim 2.4$ may be a more appropriate estimate of the nuclear attenuation in this object.

Figure~\ref{fig:mbhz} places \target\ in context among the numerous AGN selected and confirmed with \JWST, as well as UV-bright QSOs selected from ground-based surveys. 
The left panel shows black hole mass vs.~redshift, and the right panel shows black hole mass vs.~host galaxy stellar mass. 
The derived black hole mass to host stellar mass ratio is $M_{\rm BH}/M_{\star} \sim 0.1$, elevated compared to the local relation \citep{reinesRelations2015} but consistent with other \JWST-selected AGN at $z>4$. 
If the rest-UV component is indeed dominated by star formation, \target\ may represent the progenitors of the brighter LRDs with Balmer-break features indicating the presence of a moderately old stellar population \citep[e.g.][]{wangRUBIES2024}. 
More generally, \target\ is consistent with the early formation of galactic bulges in an inside-out growth paradigm \citep{roperFirst2023}. 
The consistently elevated $M_{\rm BH}/M_\star$ ratio for \JWST-selected AGN, despite a large dynamic range in mass, suggests that these objects can sustain significant simultaneous BH/galaxy growth (i.e., co-evolution) before the galaxy ``catches up'' \citep[e.g.][]{kokorevSilencing2024}.

The ubiquity of compact star-formation in the early universe is now a recurring theme with \JWST.
Many of the ultra-luminous galaxies at $z>10$ have been revealed to be very compact in the rest-UV (e.g.~GN-z11, \citealt{bunkerJADES2023}, \citealt{maiolinoSmall2024}; GHZ2, \citealt{castellanoJWST2024,zavalaDetection2024}; GN-z9p4, \citealt{schaererDiscovery2024}). 	
Moreover, gravitational lensing has allowed measurements of galaxy sizes down to tens of pc, finding ultra-compact starbursts \citep{williamsMagnified2023} with remarkable SFR surface densities $\Sigma_{\rm SFR} \gtrsim 1000$\,M$_\odot$ yr$^{-1}$ kpc$^{-2}$. 
The existence of these relatively massive, ultra-compact stellar populations may require reduced feedback efficiency at high-$z$, due to feedback-free starbursts \citep{dekelEfficient2023} or virial accelerations induced by concentrated dark matter profiles \citep{boylan-kolchinAccelerated2024}.

\begin{figure}
\centering
\includegraphics[width=\linewidth]{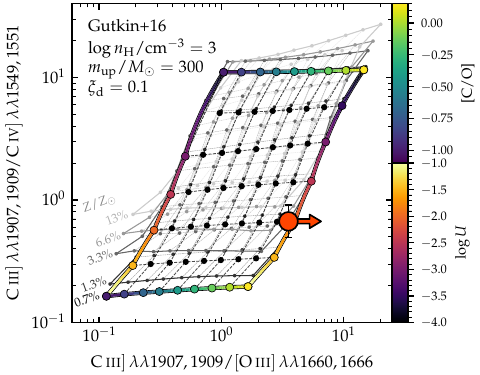}	
\caption{Rest-frame UV \ciii/\civ\ vs.~\ciii/\oiii, highlighting the potentially elevated C/O ratio in \target.  
We plot model grids for star-forming galaxies from \citet{gutkinModelling2016} at five metallicities between $0$--$20\%$ solar (metallicities $>20\%$ solar are inconsistent with the measured \ciii/\civ\ ratio).
We outline the parameter space spanned by the grids at $Z/Z_\odot \sim 0.7\%$ with colored lines indicating the varying ionization parameter $\log U$ and carbon-to-oxygen abundance ratio [C/O].
The non-detection of \oiii\,$\lambda\lambda$1660,1666, in \target\ implies a super-solar C/O, which may indicate a bursty star-formation history or exotic stellar populations, such as supermassive stars.}\label{fig:CO}
\end{figure}

Compact star formation may be associated with unique abundance patterns, particularly elevated C/O or N/O \citep{cameronNitrogen2023,harikaneJWST2024,schaererDiscovery2024,jiGANIFS2024}. 
While we do not detect UV nitrogen lines in \target, we note that the non-detection of \oiii$\,\lambda\lambda1663$ may imply significantly elevated C/O. 
Figure~\ref{fig:CO} compares the lower limit on \ciii/\oiii\ for \target\ ($\sim 3.5$) to the model grids of \citet{gutkinModelling2016}. 
We focus on models at low metallicity ($Z/Z_\odot \lesssim 20\%$), which is a reasonable assumption given the detection of these high-ionization lines. 
The model grids imply [C/O] $>0$, i.e. a super-solar C/O abundance. 
This is $\sim 0.5$ dex above the maximum observed for low-metallicity galaxies in the local universe \citep{bergChemical2019}. 
While tentative, the elevated C/O abundance may indicate a very bursty star-formation history, with carbon enrichment from the winds of AGB stars from a burst $\gtrsim 100$ Myr ago \citep{bergChemical2019,kobayashiRapid2024,hsiaoFirst2024}, or exotic stellar populations, such as supermassive stars \citep{charbonnelNenhancement2023, deugenioJADES2023}. 
Alternatively, the elevated \ciii/\oiii\ may be due to very high temperatures or densities, beyond the range of the \citet{gutkinModelling2016} models.

Finally, we note that the strong \Lya\ emission (EW $\sim 265$\,\AA) likely indicates that \target\ lives in an ionized bubble. 
This is not particularly surprising at $z=7$, past the halfway point of reionization, but the strongly 
Moreover, high \civ/\ciii\ ratios have been found to correlate with strong LyC leakage in $z\sim 3$ galaxies \citep{schaererStrong2022,kramarenkoLinking2024}; the ratio we measure for \target, $\sim 1.4$, would imply $f_{\rm esc} > 10\%$. 
Given the possible coexistence of compact/strongly ionizing star-forming galaxies and AGN, future efforts to determine the relative role of AGN/galaxies in driving reionization will require additional nuance \citep[see e.g.][]{madauCosmic2024,grazianWhat2024}.

Nevertheless, the possibility remains that the photoionization in the rest-UV is in fact dominated by the AGN. 
The UV line ratio diagnostics indeed place \target\ in the composite region, consistent with some low-$z$ Type I AGN, and the lack of significant \mgii\ emission could be due to low BLR metallicity \citep{shinStrong2021,wangMetallicity2022} or related to the Eddington ratio and covering factor \citep{dongEddington2009}. 
Future deeper and/or higher-resolution observations (e.g. with NIRSpec G140M/G235M) may be able to detect \heiiAuv\ or \mgiiA, or identify any broad components in the rest-UV emission lines, helping to better disentangle the AGN and host galaxy components.  
At the same time, larger spectroscopic samples of LRDs will be needed to better constrain the abundance of compact galaxy components.

\facilities{\textit{JWST} (NIRCam, NIRSpec). The \JWST\ data used in this work can be found in MAST: \dataset[10.17909/sb0f-gb28]{https://dx.doi.org/10.17909/sb0f-gb28}.}

\software{\texttt{astropy} \citep{astropycollaborationAstropy2013}, \texttt{matplotlib} \citep{hunterMatplotlib2007}, \texttt{numpy} \citep{harrisArray2020}, STScI JWST Calibration Pipeline \citep[\url{jwst-pipeline.readthedocs.io};][]{rigbyScience2023}.}

\section*{Acknowledgements}

H.B.A. acknowledges the support of the UT Austin Astronomy Department and the UT Austin College of Natural Sciences through Harrington Graduate Fellowship, as well as the National Science Foundation for support through the NSF Graduate Research Fellowship Program.
C.M.C. thanks the National Science Foundation for support through grants AST-2009577 and AST-2307006 and to NASA through grant JWST-GO-01727 awarded by the Space Telescope Science Institute, which is operated by the Association of Universities for Research in Astronomy, Inc., under NASA contract NAS 5-26555. 

Authors from UT-Austin acknowledge that they work at an institution that sits on indigenous land. 
The Tonkawa lived in central Texas, and the Comanche and Apache moved through this area. 
We pay our respects to all the American Indian and Indigenous Peoples and communities who have been or have become a part of these lands and territories in Texas.

\newpage

\newpage
\bibliographystyle{aasjournal} 
\bibliography{LRD_UV_paper.bib}

\end{document}